\documentstyle[amstex,aps,prl,multicol,epsfig,amssymb]{revtex}
\begin{document}
\draft
\title{Ginzburg-Landau theory and effects of pressure 
on a two-band superconductor : application
to MgB$_2$}
\author{J.~ J. Betouras, V.~A. Ivanov\cite{valeri}, F.~M. Peeters}
\address{
Department of Physics,
University of Antwerpen (UIA),
Universiteitsplein 1,
B-2610 Antwerpen,
Belgium}
\date{\today}
\maketitle

\begin{abstract}
We present a model of pressure effects 
of a two-band superconductor based on a Ginzburg-Landau free energy
with two order parameters.
The parameters of the theory are pressure
as well as temperature dependent. New pressure effects emerge as a result
of the competition between the two bands. The theory then is applied to
MgB$_2$. 
We identify two possible scenaria regarding the fate of the two $\sigma$
subbands under pressure, depending on whether or not
both subbands are above the Fermi energy at ambient pressure. 
The splitting of the two subbands is probably caused by the
$E_{2g}$ distortion. If only one subband is above the Fermi energy
at ambient pressure (scenario I), 
application of pressure diminishes the splitting and it is possible
that the lower subband participates in the superconductivity.
The corresponding crossover pressure and
Gr$\ddot{u}$neisen parameter are estimated. In the second scenario
both bands start above the Fermi energy and they move below it,
either by pressure or via the substitution of Mg by Al. In both scenaria,
the possibility of electronical topological transition is emphasized.
Experimental signatures of
both scenaria are presented and
existing experiments are discussed in the 
light of the different physical pictures.

\end{abstract}

\begin{multicols}{2}
\section{Introduction}
The discovery of superconductivity in the material MgB$_2$ 
\cite{nagamatsu} initiated intensive recent theoretical and experimental 
interest. 
The possibility 
of a high critical temperature in a class of materials which are chemically 
much simpler than the high-T$_c$ cuprates 
and the occurence of large critical current 
densities pose some interesting new questions in the 
research of superconductivity. 
MgB$_2$ is a type II superconductor in the clean limit 
\cite{finnemore}.
The crystal belongs to the space group $P6/mmm$ or $AlB_2$-structure 
where borons are packed in honeycomb layers alternating with hexagonal layers 
of magnesium ions.
The ions Mg$^{2+}$ are positioned above the centers of hexagons formed by boron
sites and donate their electrons to the boron planes.
The electronic structure is organized by the narrow energy bands with near
two-fold degenerate ($\sigma$-electrons) and the wide-band ($\pi$-electrons)
\cite{armstrong,medvedeva,kortus}. 
Without any of the
lattice strain, the $\sigma$ dispersion relations are slightly splitted 
due to the two boron atoms per unit cell (the electronic analog of Davydov
splitting for phonons). The $\sigma$ portions of the
Fermi surface (FS) consist of coaxial cylinders along the $\Gamma$ - 
A symmetry direction of the Brillouin zone (BZ), whereas the $\pi$- bands are strongly dispersive. 

The excited phonon modes in MgB$_2$ present a sharp cut-off at about 100 meV.
The optical modes ($B_{1g}$, $E_{2g}$, $A_{2u}$ and
$E_{1u}$) are practically non-dispersive, along the $\Gamma - A$  
direction. 
Various Raman data \cite{goncharov,struzhkin,hlinka} show a small 
spread of the $E_{2g}$ frequency around 74.5 meV in different ceramic samples
of MgB$_2$. 
The strong deformation potential $\Delta$ of the in-plane $E_{2g}$ 
mode \cite{an,yildirim} causes a significant energy splitting
of the $\sigma$ band around 1.5 eV, lifting its two-fold degeneracy at
the $\Gamma$-point of the BZ.
In general, measurements of the effect of pressure on the electronic 
structure are based
on the Raman technique, therefore only the influence on the $E_{2g}$ mode can
be traced. High pressure experiments up to 15 GPa 
\cite{goncharov} have revealed a large increase of the out of phase $E_{2g}$
phonon mode. This leads to a suppression of the displacement of the boron
atoms  as well as of the deformation potential $\Delta = B_2 u^2 + B_4 u^4$
where {\it u} is the displacement \cite{yildirim} which will be used below.
Moreover, experiments under pressure up to 40 GPa do not show
any structural phase transition \cite{goncharov,prassides,bordet,monteverde}.
Due to the anisotropy the hydrostatic pressure decreases the 
inter-plane distance
more than the in-plane distance between adjacent borons.

The interpretation of experimental data from different spectroscopic methods 
\cite{chen,tsuda,szabo},  
suggests    
the presence of two different superconducting gaps.
The specific heat behavior \cite{junod}, the low isotope effect
\cite{budko}, pressure effects \cite{schilling} and penetration
depth data \cite{panagopoulos} 
provide evidence for a complicated superconducting
order parameter in MgB$_2$.
Recent NMR measurements \cite{pissas} of $^{11}B$ as well as Hall 
measurements \cite{jin} are consistent with 
a $\sigma$-band driven superconductivity in MgB$_2$ where the $\pi$ band
participates due to interband scattering. 
The use of two order parameters is justified by recent
experimental results on scanning tunneling microscopy \cite{giubileo}.

Some microscopic theories \cite{an,theory1} are based on $\sigma$ band
or $\sigma-\pi$ band scenaria for superconductivity whereas others
\cite{theory2} are concentrated on a $\pi$ band superconductivity.
A consensus is formed though, that the driving band for the
superconductivity is the $\sigma$ one.
The calculated spectral functions \cite{knigavko} and the analysis of the
reflectance measurements \cite{marsiglio} show the possibility of
different superconducting mechanisms beyond the conventional
electron-phonon Bardeen-Cooper-Schrieffer (BCS) pairing. 
The complicated nature of superconductivity
in MgB$_2$ does not allow at present to accept the conventional
superconducting mechanism. Therefore, it is appropriate 
to construct a phenomenological
theory.
 
Since MgB$_2$ presents an example of a two-band superconductor
\cite{theory3} may serve as a paradigm to investigate new effects.
The purpose of this article is to present a 
phenomenological Ginzburg-Landau (GL) theory for a two-band superconductor 
appropriate to MgB$_2$ based on two order parameters and, in that framework, 
to provide a physical picture of pressure 
effects on MgB$_2$. The study of pressure effects
is one way to investigate (i) the different topology of the two bands and
(ii) the question of the participation of the two $\sigma$ subbands
in the superconductivity. A brief account of a part of this work
was presented elsewhere \cite{leshouches}. In the present article
missing terms in the GL functional are restored, non-adiabatic effects
are explained using a different 
approach and 
the effects of pressure are discussed in more detail 
and compared with very recent experiments.
The work is organized as follows. In section II  the
GL description is presented, in Section III the pressure effects, in
connection with the $\sigma$ band, are discussed and a general
discussion with some concluding remarks are given in Section IV.

\section{Ginzburg-Landau description}

We introduce the GL free energy functional with two order parameters,
appropriate for MgB$_2$, assuming that both the order parameters
belong to the $\Gamma_1$ representation of the point group of MgB$_2$ crystal. 
The two order parameters
are labeled by the corresponding band ($\sigma$ or $\pi$) 
of MgB$_2$, without loss of generality :
\end{multicols}
\begin{eqnarray}
\nonumber
F = \int d^3r \{ \frac{1}{2m_{\sigma}} |\vec{\Pi} \psi_{\sigma}|^2 &+& 
\alpha_{\sigma} |\psi_{\sigma}|^2 + \beta_{\sigma} |\psi_{\sigma}|^4 + 
\frac{1}{2m_{\pi}} 
|\vec{\Pi} \psi_{\pi}|^2 + \alpha_{\pi} |\psi_{\pi}|^2 + 
\beta_{\pi} |\psi_{\pi}|^4 + r (\psi_{\sigma}^* \psi_{\pi} + 
\psi_{\sigma} \psi_{\pi}^*) \\
&+& {\gamma}_1 ({\Pi}_x {\psi}_{\sigma} {\Pi}_x^* {\psi}_{\pi}^*
+ {\Pi}_y {\psi}_{\sigma} {\Pi}_y^* {\psi}_{\pi}^* + c.c.)+ 
\beta |\psi_{\sigma}|^2 |\psi_{\pi}|^2\ +
\frac{1}{8 \pi} (\vec{\nabla} \times \vec{A})^2 \},
\end{eqnarray}
\begin{multicols}{2}
\noindent where $\vec{\Pi} = -i \hbar \vec{\nabla} - 2e/c \vec{A}$ 
is the momentum operator, $\vec{A}$ 
is the vector potential and $\alpha_{\sigma,\pi} = \alpha_{\sigma,\pi}^0 
(T-T_{c\sigma,\pi}^0)$. It is possible to take into account the large 
anisotropy of the two order parameters (the $\sigma$ is almost
two-dimensional) by  rescaling of the axis according to the effective
masses (directional dependence effective mass). In the present work, since we
are focusing exclusively on pressure effects, the derivative terms are
not important.

In Eq. (1) we have used the fact that the mixing $r$-term favors the 
coupling of linear
combination of the two order parameters with phase difference 0 if $r<0$ 
or $\pi$ if $r >0$ 
between them \cite{betouras},
therefore a term $\psi_{\sigma}^{2*}\psi_{\pi}^2 + c.c.$ is 
incorporated into the 
$\beta$-term of the free energy \cite{note1}.
If $r = 0$ then Eq. (1) is the free energy for two bands 
without Josephson coupling between them. The quartic $\beta$-term is the
only one which mixes the two gaps and is unimportant in this case.
The onset of the superconducting state in one band does not imply the onset
in the other. This corresponds to the case of $V_{sd}=0$ in the two-band
BCS treatment \cite{suhl}.
If $r \neq 0$, which is the case of MgB$_2$,
then the pair transfering term is present in the GL functional 
and it means 
that the onset
of superconductivity in one band implies automatically the appearance of
superconductivity in the other. There is a single observed T$_c$ which is
a function of the bare ones T$_{c\sigma,\pi}^{0}$.
In MgB$_{2}$ the compression due to pressure 
is anisotropic\cite{goncharov,tissen,vogt}.
According to Ref.~%
\onlinecite{goncharov} the compressibility along the $c$-axis is almost twice
larger than that of the plane compressibility.
Therefore application of uniaxial pressure on single crystals 
will affect differently the two gaps.
Following Ozaki's formulation
\cite{ozaki,sigrist}, we may add to the GL functional, Eq. (1), the 
term which couples the order parameters, in second 
order,  with the strain tensor $\epsilon$ to first order, having
already specified the symmetry of the order parameters : 
\begin{eqnarray}
\nonumber
F_{strain} = &-& C_1(\Gamma_1) [\delta (\epsilon_{xx}+\epsilon_{yy}) +
 \epsilon_{zz} ] |\psi_{\sigma}|^2 \\
\nonumber
&-& C_2(\Gamma_1) (\epsilon_{xx}+\epsilon_{yy} +
 \epsilon_{zz}) |\psi_{\pi}|^2 \\
&-& C_3(\Gamma_1) [\delta' (\epsilon_{xx}+\epsilon_{yy}) +
 \epsilon_{zz} ] (\psi_{\sigma}^* \psi_{\pi} + c.c.), 
\end{eqnarray}
where $C_{1,2,3}(\Gamma_1)$ are coupling constants and 
$\delta$, $\delta'$ are given in terms of
the elastic constants of the material. 
This will lead to a change of the bare critical temperatures
$T_{c\sigma,\pi}^{0}$ and consequently of the actual critical temperature
$T_c$. The physical reasoning behind this is that the material 
will change in such a way as to gain condensation energy by
enhancing the density of states in the direction where the superconducting
gap is larger. The term proportional to $\psi_{\sigma}^* \psi_{\pi}+ c.c.$
in the $F_{strain}$ will renormalize the mixing coupling $r$ to $\tilde{r}$.
We model the
corresponding differences by taking a linear dependence of the two bare
$T_c$'s with pressure $T_{c\sigma,\pi} =T_{c\sigma,\pi}^0 - \eta_{\sigma,\pi}p$
, where $\eta_{\sigma,\pi} =|\partial T_{c\sigma,\pi}^0(p=0)/\partial p|$
and we will discuss the validity of the assumption for the MgB$_2$ later.

Analyzing the equations which result
from the minimization of the GL free energy we get : 
\begin{equation}
\alpha_{\pi} \alpha_{\sigma}  =  {\tilde{r}}^2.
\end{equation}
This gives the pressure dependence of the superconducting critical temperature:
\begin{eqnarray}
\nonumber
T_c(p) &=& \frac{1}{2} \left[ T_{c\sigma}^0 + T_{c\pi}^0 - 
(\eta_{\pi}+\eta_{\sigma})p \right]   \\
&+& \frac{1}{2} 
\{[T_{c\sigma}^0 - T_{c\pi}^0 + (\eta_{\pi} - \eta_{\sigma})p]^2 + 
a^2 \}^{1/2},
\end{eqnarray}
where $a^2 = 4 {\tilde{r}}^2/
(\alpha_{\pi}^0 \alpha_{\sigma}^0)$. Deviations from a straight line at 
moderate values of pressure can be attributed to the two bands.
From the above formula the inequality $dT_c/dp < 0$ is always true 
as long as the renormalized mixing parameter $\tilde{r}$ is real,
as a consequence of the initial assumption on the pressure dependence
of the bare critical temperatures. 
In the case of MgB$_2$ 
we can safely consider that $\eta_{\pi}>\eta_{\sigma}$.

\section{Nonadiabatic effects}

We now wish to address an issue  specific to MgB$_2$  which can be also
realized in other multiband superconductors.
The physical situation we would like to question is the splitting of
the two $\sigma$ subbands at ambient pressure 
and their participation in superconductivity with the increase of pressure.
Due to contradictory experimental as well as theoretical results we proceed
by distinguishing two different scenaria. The first one addresses 
the splitting of the two sigma subbands at ambient pressure
as shown by first princible calculations
\cite{an,yildirim,boeri} when one of the subbands is below the
Fermi energy $E_F$. The second scenario addresses the opposite situation
where both the $\sigma$ subbands are above $E_F$ at ambient pressure. 

\subsection{Scenario I : one $\sigma$ sub-band above E$_F$ at 
ambient pressure.}

In particular in \cite{boeri} the splitting of the two subbands
was studied in detail and compared with similar situation in AlB$_2$.
The conclusion is that due to the $E_{2g}$ phonon mode, the two $\sigma$
bands split nearby $\Gamma$ point and the lower band completely 
sinks below the Fermi
energy. This is the case for MgB$_2$ and also for
the heavily hole-doped graphite. Moreover, experimental results 
\cite{tissen,meletov} showed a kink in the superconducting 
critical temperature 
as a function of 
pressure at approximately 6-8 GPa and also a kink in the volume
dependence of $T_c$ for Mg$^{10}$B$_2$ at around 20 GPa and
Mg$^{11}$B$_2$ at around 15 GPa \cite{struzhkin}.

The physics we discuss here, is the change in the 
electronic properties of the material under pressure and its 
influence on the superconducting state. 
The band which is below the Fermi
level at ambient pressure \cite{an} is possible to overcome
the energy difference and to get above the
Fermi level at a certain value of pressure (termed as crossover pressure), 
restoring the degeneracy of the two $\sigma$ bands at point $\Gamma$. 
This is a  non-adiabatic effect as discussed in \cite{boeri}.
To understand and illustrate this effect on the superconducting state
we need to consider the pressure
dependence of the coefficient $\alpha_{\sigma}$ as usual \cite{bob} 
with the modification due to the particular physical
situation by writing:
\begin{equation}
\alpha_{\sigma}= \alpha_{\sigma}^{0} (T-T_{c\sigma}^{0}+ \eta_{\sigma} p)
+\alpha_{\sigma}^{1}(p-p_c) \Theta(p-p_c), 
\end{equation}
\noindent where $\Theta(x)$ is the Heaviside step function, 
$\alpha_{\sigma}^{0}$ and $\alpha_{\sigma}^{1}$ are positive constants.
This choice reflects the fact that at a certain crossover
pressure $p_c$ the second band 
starts to participate in superconductivity
as well and that initially it is an increasing function of pressure.
We obtained the above
formula by requiring the continuity of the coefficient $\alpha_{\sigma}$
at $p_c$.

After the energy difference between the two $\sigma$ bands almost disappears,
then they both follow the same reduction of the bare $T_c$'s with pressure.
The approximate value of the  crossover pressure 
can be estimated as follows. 
The energy difference  between the two subbands is approximated by 
$ \delta E \simeq%
\left( 1-n\right)  \sqrt{\Delta} $, where the fraction of
superconducting electrons is $1-n\sim 0.03$ \cite{armstrong}, 
{\it i.e.} the carrier
density per boron atom in Mg$B_{2}$. The approximate value of 
$p_{c}$ is half the value of the pressure which
 suppresses the deformation potential 
$\Delta $. 
The deformation potential in turn, can be estimated by the expression 
$p_{c}\Omega \sim \left( 1-n \right) \sqrt{\Delta }$
where $\Omega \sim 30$\AA $^{3}$ is a unit cell volume of Mg$B_{2}$ and $%
\Delta =0.04eV^{2}$ is the deformation potential for a boron displacement $%
u\sim 0.03$\AA\ \cite{yildirim}. Using these parameters, we get a
crossover pressure of $p_c \sim 15 GPa$. This estimate shows, 
that a realistic applied pressure
influences drastically the electronic structure and the FS topology, 
restoring the
degeneracy of the two subbands at the $\Gamma$-point which are 
initially splitted.
Superconductivity in the second subband may occur at lower 
values of the estimated $p_c$ due to the fact that both subbands are affected. 
The elimination of the energy difference 
around the Fermi energy will also occur at lower values due to the 
corrugation of the $\sigma$ portions of the 
FS, the already existing strains in the
material and the anisotropic compressibility. These considerations 
make the above estimate an upper limit of the crossover pressure $p_c$.
The degree to which shear stresses of the sample and the
surrounding fluid under pressure affect the data of
pressure measurements is still under investigation \cite{schilling}.

The microscopic nature of this crossover is caused by the change of the
FS. If the topology of the FS changes under pressure (this subsection)
or due to substitutions
in the composition (next subsection), 
a van Hove singularity at critical energy $E_c$ in the electronic
density of states $\rho(E)$ is manifested in an electronic topological
transition of the 5/2 kind \cite{lifshits}. Since 
$dT_c/dp$ is essentially proportional to the derivative $d\rho(E_F)/dE_F$,
it develops singularities $\propto 1/\sqrt{E_F-E_c}$, provided that
the strength of carrier attraction varies slightly with pressure
\cite{makarov}.

In Fig. 1 using a model calculation, 
we illustrate
the expected behavior of $T_c$ and  
the form of the order parameter as a function of pressure
at fixed temperature and, schematically, the kink of $T_c$ at $p_c$
due to the change of the slope from Eq. (5). 
The predicted kink close to $p_c$ can be detected directly in 
a penetration depth experiment under pressure. The chosen parameters
are such that they respect the relation $\eta_{\pi} > \eta_{\sigma}$,
the fact that the superconducting density coming from the $\sigma$
band is higher than the one coming from the $\pi$ band
($\alpha_{\sigma} > \alpha_{\pi}$) and that the driving band
for superconductivity is $\sigma$ ($T_{\sigma}^0 > T_{\pi}^0$). 

The contribution on the Gr$\ddot{u}$neisen parameter :
\begin{equation}
G= B\frac{dln\omega}{dp} =\frac{B}{\omega} \frac{d\omega}{dp},
\end{equation}

\noindent where B is the bulk modulus and $\omega$ is the phonon 
frequency, can be
approximately obtained if we observe that $\delta E \simeq \hbar \omega_{E2g}$
then $G \simeq  B/2p_c$. The bulk modulus is measured to be B$\simeq
114$ GPa \cite{li} and our estimate for the Gr$\ddot{u}$neisen parameter is 
G$ \simeq $ 3.8.
Experimentally G$= 2.9 \pm 0.3$ as reported in \cite{goncharov}
for the measured Raman active E$_{2g}$ phonon mode. 
However, as indicated in Ref. \cite{goncharov}, for anisotropic crystals it
would be more appropriate to scale the frequency shift in the
Gr$\ddot{u}$neisen parameter with the variation of the lattice constant $a$
such as $G =dln\omega/ 3 dlna$ \cite{hamfland}. Then, the Gr$\ddot{u}$neisen
parameter for the MgB$_2$ takes the value 3.9 $\pm$ 0.4.
There is an excellent agreement with the above estimate and it justifies
our approach.

\begin{minipage}[b]{.99\linewidth}
\begin{figure}
\begin{center}
\epsfig{figure=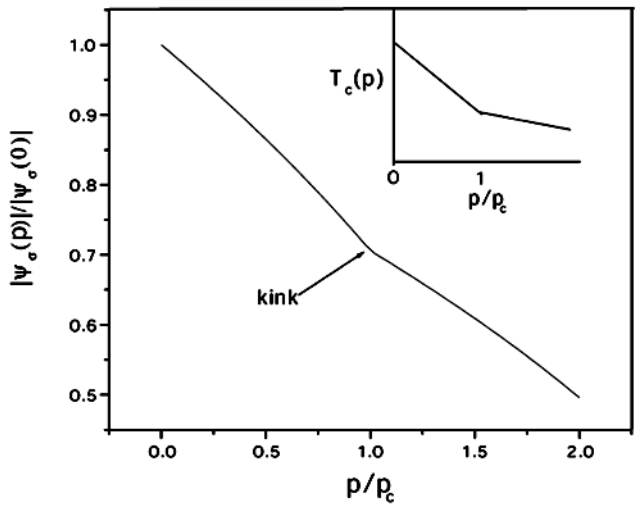,height=7.0cm}
\end{center}
\caption{The form of the order parameter 
as a function of pressure, where the kink at
p$_c$ is shown. Inset: the behavior of the critical temperature
T$_{c}$(p). The chosen parameters are (in
reduced units):   T=0.8 T$_{c\sigma}^0$,
T$_{c\pi}^0$=0.7 T$_{c\sigma}^0$, $\eta_{\sigma}$= 0.2 T$_{c\sigma}^0$/p$_c$,
$\eta_{\pi}$= 0.5 T$_{c\sigma}^0$/p$_c$, $\alpha_{\sigma}^0$=2,
$\alpha_{\pi}^0$=1, $\alpha_{\sigma}^1$= 0.1.}
\label{fig:fig1}
\end{figure}
\end{minipage}

\subsection{Scenario II : both $\sigma$ sub-bands above E$_F$ at 
ambient pressure.}

Some recent experimental and theoretical studies
\cite{postorino} lend support to a second scenario.
More specifically, the substitution of Mg by the higher valent 
Al was used to
provide the necessary change of the Fermi energy via electron
concentration and the lattice parameters in the compound
Mg$_{1-x}$Al$_x$B$_2$.
This is in essence equivalent to a change of Fermi surface topology
as in the pressure experiments.
As it was shown in \cite{postorino} by increasing the Al content
(equivalently by increasing the pressure), a topological crossover
occurs at some point and the negative slope of $T_c$ as a function of
Al content $x$ decreases further giving a different picture
from that of Fig. 1.

To model this in the language of the GL model, one basically has to
substitute the pressure p and the crossover pressure 
$p_c$ with the Al concentration $x$ and and the crossover concentration
$x_c$ respectively, requiring again continuity at $x_c$:
\begin{equation}
\alpha_{\sigma}= \alpha_{\sigma,x}^{0} (T-T_{c\sigma}^{0}+ \eta_{\sigma,x} x)
+\alpha_{\sigma,x}^{1}(x-x_c) \Theta(x-x_c) 
\end{equation}

\noindent where the extra substrcipt $x$ denotes that the numerical 
coefficients
are different from Eq. (5). The critical temperature as a function of the
Al content is illustrated in Fig. 2. 
The physics is similar to the topological transition
described in scenario I where now instead of pressure, the changing
parameter is the composition. It is a more controlled way to change the
topology of the FS.
In fact, an abrupt topological change in the $\sigma$-band Fermi surface 
was found at $x = 0.3$ in Ref. \cite{postorino}.
When the $\sigma$ bands are filled in Mg$_{1-x}$Al$_x$B$_2$ at 
$x \approx 0.6$, superconductivity dissapears. It is worth noticing
that in this compound, the impurity scattering broadens the
van Hove singularities at the saddle points E$_c$ of the FS
and smears the change of slope of $T_c$ at E$_c$. Therefore
the kink of $T_c$ in Fig. 2 will be less pronounced
(see also the experiment in Ref. \cite{postorino}).

We emphasize here the need of de Haas-van Alphen (dHvA) data for 
Mg$_{1-x}$Al$_x$B$_2$, especially close to x=0.3. The pressure
derivative of the superconducting 
critical temperature as a function of the concentration $dT_c(x)/dp$
can be also conclusive since it can clarify the correlation between
pressure, Al concentration and the kink at $x_c$.

\begin{minipage}[b]{.99\linewidth}
\begin{figure}
\begin{center}
\epsfig{figure=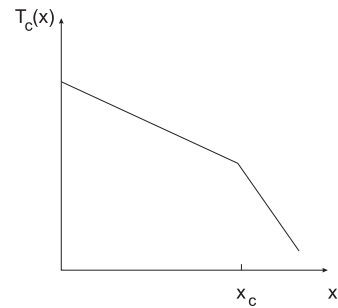,height=4.0cm}
\end{center}
\caption{Schematically the form of the superconducting critical temperature 
T$_c$(x) in the second scenario. x is the
concentration of Al in a Mg$_{1-x}$Al$_x$B$_2$-like system.}
\label{fig:fig2}
\end{figure}
\end{minipage}

\section{Discussion}

Interestingly enough, there is an obvious discrepancy from the two
pictures of the previous section, which calls for more experimental data.
Observation of quantum oscillations in dHvA
experiments provides
the information about the bands, the effective mass of the carriers
and the shape of the Fermi surface in MgB$_2$. In Ref. \cite{yelland},
$only$ the smaller cylindrical Fermi surface along the c-axis
was observed
(with a hole band mass -0.251 m$_e$, where m$_e$ is the electron 
mass).
The same authors reported the observation of $only$ the larger of the 
two $\sigma$ tube-like Fermi surfaces (with a band mass -0.553 m$_e$)
in Ref. \cite{carrington}. Although the estimated parameters of the MgB$_2$
are in agreement with the LDA band calculations \cite{lda} future dHvA
experiments should clarify the detailed observation 
of the $\sigma$ portions of the FS.
Also note that angle resolved photoemission spectroscopy 
experiments \cite{uchiyama}
detected $one$ band along the $\Gamma-K$ symmetry line of the BZ.

There are a few points in order to discuss further with respect 
to applicability of the theory developed here.
Early pressure experiments \cite{monteverde} demonstrated the overall 
decrease of $T_c$ with a pressure increase which was attributed
to the loss of holes. In two of the samples of \cite{monteverde} 
there is a linear dependence
of $T_c$ on pressure and in two others a weak quadratic dependence.
We stress that the samples were polycrystalline and the experiment is 
effectively under hydrostatic pressure. Also the degree of nonstoichiometry 
was not known. The almost linear dependence for a wide range of 
pressures, makes the GL functional as presented, valid for MgB$_2$.

Experiments on single crystals
will be able to clarify the effects which are described.
We do not attempt at the moment any actual detailed 
fitting of experimental data,
except from the illustrative fitting in section III,
because there is no experimental consensus on the different values of 
key parameters of the theory ({\it e.g.}, 
there is a wide range of published data
on the value of $dT_c/dp$ \cite{schilling}).                                   More detailed Raman data with optical reflectivity, specific 
heat and thermal conductivity measurements under pressure in the 
superconducting phase 
are needed as well as a penetration depth experiment
which can reveal the pressure dependence of the superfluid density.
Also the experiments suggested in scenaria I and II will be crucial in
the question regarding the topological transition, where $E_F$ can be varied
experimentally either by external pressure (Sec. IIIA) or by alloying
(Sec.IIIB).
Recent Raman spectra at high pressure \cite{meletov} 
reveal a reduction of the slope of the pressure-induced frequency shift by
about a factor of two, at about 18 GPa which supports the suggestion that
MgB$_2$ may undergo a pressure-induced topological electronic transition.
In connection with the appearance of superconductivity under pressure,
we stress the high pressure superconducting phase of CaSi$_2$ which
superconducts above an applied pressure of 12 GPa with $T_c \sim  14 K$
\cite{sanfilipo}. 

One proposed experiment which can potentially show the different role 
of the two bands ($\sigma$ and $\pi$) is the detection of the 
Leggett mode or the internal Josephson current \cite{agterberg}. 
Moreover an upward curvature in H$_{c2}$ is also known to be a signature
of multicomponent superconductivity.

In summary, we present an analysis of pressure effects, within
the GL theory,  of a two-band
superconductor and apply it to 
MgB$_2$. We
make predictions for non-adiabatic effects
and discuss different experiments from which 
crucial information can be extracted on the physics of the
two participating bands $\sigma$ and $\pi$ as well as the 
more delicate questions on the two $\sigma$ subbands.

We are grateful to Daniel Agterberg, Christos Panagopoulos, Robert Joynt,
Manfred Sigrist, Stefan-L$\ddot{u}$dwig Drechsler
and the anonymous referee for very useful comments and
constructive criticisms.
This work was supported by the Flemish Science Foundation (FWO-Vl), the
Inter-University Attraction Poles Programme (IUAP) and the University of
Antwerpen (UIA).


\end{multicols}


\begin{thebibliography}{99}
\bibitem{note1} The difference in the form of the free energy with a form
derived for a possible $s$ and $d$ mixing in high-$T_c$ superconductors is
in the mixed derivative term $\gamma_1$, where we add the contributions 
from the two directions instead of subtracting them.
This reflects the fact that a $d$-wave order parameter under
$90^0$ rotation changes sign whereas here the assumed 
symmetry is $\Gamma_1$ for both order parameters.
\end{thebibliography}
\end{document}